\pdfoutput=1

\documentclass[acmsmall, manuscript, screen, review=False]{acmart}

\usepackage{wrapfig}
\usepackage{subcaption}

\setcopyright{none}
\makeatletter
\renewcommand\@formatdoi[1]{}
\makeatother

\acmConference[CSCW '24]{The 27th ACM SIGCHI Conference on Computer-Supported Cooperative Work \& Social Computing (CSCW)}{November 9-13, 2024}{San Jose, Costa Rica}

\title[AudienceView]{AudienceView: AI-Assisted Interpretation of Audience Feedback in Journalism}

\author{William Brannon}
\email{wbrannon@mit.edu}
\orcid{0000-0002-1435-8535}
\affiliation{
  \institution{MIT Center for Constructive Communication \& MIT Media Lab}
  \city{Cambridge}
  \state{MA}
  \country{USA}
}

\author{Doug Beeferman}
\email{dougb5@mit.edu}
\orcid{0009-0005-5879-5744}
\affiliation{
  \institution{MIT Center for Constructive Communication \& MIT Media Lab}
  \city{Cambridge}
  \state{MA}
  \country{USA}
}

\author{Hang Jiang}
\email{hjian42@mit.edu}
\affiliation{
  \institution{MIT Center for Constructive Communication \& MIT Media Lab}
  \city{Cambridge}
  \state{MA}
  \country{USA}
}

\author{Andrew Heyward}
\email{aheyward@media.mit.edu}
\affiliation{
  \institution{MIT Center for Constructive Communication \& MIT Media Lab}
  \city{Cambridge}
  \state{MA}
  \country{USA}
}

\author{Deb Roy}
\email{dkroy@mit.edu}
\orcid{0000-0002-2780-4768}
\affiliation{
  \institution{MIT Center for Constructive Communication \& MIT Media Lab}
  \city{Cambridge}
  \state{MA}
  \country{USA}
}



\begin{document}

\begin{abstract}
Understanding and making use of audience feedback is important but difficult for journalists, who now face an impractically large volume of audience comments online. We introduce AudienceView, an online tool to help journalists categorize and interpret this feedback by leveraging large language models (LLMs). AudienceView identifies themes and topics, connects them back to specific comments, provides ways to visualize the sentiment and distribution of the comments, and helps users develop ideas for subsequent reporting projects. We consider how such tools can be useful in a journalist's workflow, and emphasize the importance of contextual awareness and human judgment.
\end{abstract}

\maketitle

\section{Introduction}
\label{sec:introduction}
Journalists have a longstanding interest in the identity of the audience and how that audience responds to the journalists' work \cite{robinsonAudienceMindEye2019}, though this impulse is balanced against the need to preserve journalistic independence. Journalism's traditional means of obtaining and considering audience feedback, however, have been upended in recent years by the rise of the Internet, which provides far more input from readers and viewers than ever before. In place of letters to the editor and Nielsen audience figures, journalists now have access to detailed clickstream information, engagement statistics from social media, and above all a wealth of reader comments. Leveraging recent advances in natural language processing, themselves driven by the same increase in the scale of web data, we aim to simplify the task of engaging with these comments and deriving actionable insights from them.

This paper introduces AudienceView,\footnote{A deployed version of the tool is available at \url{https://frontline.ccc-mit.org/}, and the underlying code is available \href{https://github.com/mit-ccc/AudienceView-demo}{on Github}.} an AI-assisted tool for journalists to make sense of audience feedback on YouTube. We have developed AudienceView in partnership with PBS's Frontline team, and development is thus targeted toward YouTube-hosted video journalism. The tool is, however, modular and extensible to other comment sources. Our evaluation applies it to all Frontline documentaries during the more than 10 years between August 2013 and January 2024 (250 in total), comprising just over 599,000 comments.

AudienceView complements prior work on AI for qualitative analysis and sensemaking \cite{beefermanFeedbackMapToolMaking2023, overneySenseMateAccessibleBeginnerFriendly2024, goldmanQuADDeepLearningAssisted2022} in the particular context of journalism. Some of its components can also be viewed as a kind of summarization, building on a long literature in NLP on that question. Because of the diversity of the comments and use cases, we have designed AudienceView with two goals in mind: a) providing more than one way to visualize and understand the data, and b) keeping a close connection to the underlying comments, and surfacing them wherever possible. Both are intended to complement, rather than replace, human judgment in producing news.

\section{Related Work}
\label{sec:related-work}

\paragraph{NLP for sensemaking and qualitative analysis}
Both commercial \cite{verbisoftwareMAXQDA2022, lumiveroNVivo2023} and academic \cite{overneySenseMateAccessibleBeginnerFriendly2024} tools use AI for qualitative analysis on textual data. These tools, however, are sometimes expensive and remain within a qualitative analysis paradigm that requires considerable effort from the analyst. A number of NLP tools and topic detection algorithms have been developed to allow more automated sensemaking from textual data. In addition to classic methods like Latent Dirichlet allocation \cite{bleiLatentDirichletAllocation2003}, more recent neural network-based techniques include top2vec \cite{angelovTop2VecDistributedRepresentations2020} and BERTopic \cite{grootendorstBERTopicNeuralTopic2022}, often based on the general idea of clustering dimension-reduced embeddings from a language model. While these methods are powerful, they require technical knowledge and are inaccessible to non-technical users. Generative AI tools like ChatGPT are more accessible, but have issues for journalistic sensemaking, especially a lack of integration with news and social sites and limited context-window lengths. We can help fill this gap, inspired by similar efforts in the domain of survey responses \cite{beefermanFeedbackMapToolMaking2023}, with a journalism-focused tool that leverages these advanced NLP methods while also handling collection of audience comments and providing a user-friendly interface.

\textit{Digital journalism}.
The question of audience understanding has received much attention in journalism and journalism scholarship over the years \cite{robinsonAudienceMindEye2019}. Journalists' ability to acquire audience feedback at all is partly a function of technology, and doing so has become much easier in the digital-media era. Social media in particular has become central to, and sped up, journalistic practice \cite{lawrenceCampaignNewsTime2020}, and much work also argues that these changes amount to ``ambient'' journalism \cite{hermidaSocialJournalismExploring2012} or ``context collapse'' \cite{marwickTweetHonestlyTweet2011} that brings the audience more deeply into the process and inspires changes in journalists' presentation of the news. Reporters and editors now put much more effort into understanding the audience \cite{whippleQualityQuantityPolicy2018}, especially via clickstream or traffic-tracking tools like Chartbeat \cite{chartbeatinc.Chartbeat2024}. Comments, while important, often receive less attention because of the unpleasant nature of abuse and trolls \cite{robinsonAudienceMindEye2019}.
On the AI side, most work on AI in journalism has considered AI as a direct producer of news articles \cite{strayMakingArtificialIntelligence2019}, or AI in journalism education \cite{pavlikCollaboratingChatGPTConsidering2023}. We aim rather to use AI for sense-making, helping journalists understand audience views.

\section{System Overview}
\label{sec:system-overview}
\begin{figure}[ht!]
    \centering
    \begin{subfigure}[b]{0.45\textwidth}
        \centering
        \includegraphics[width=\textwidth]{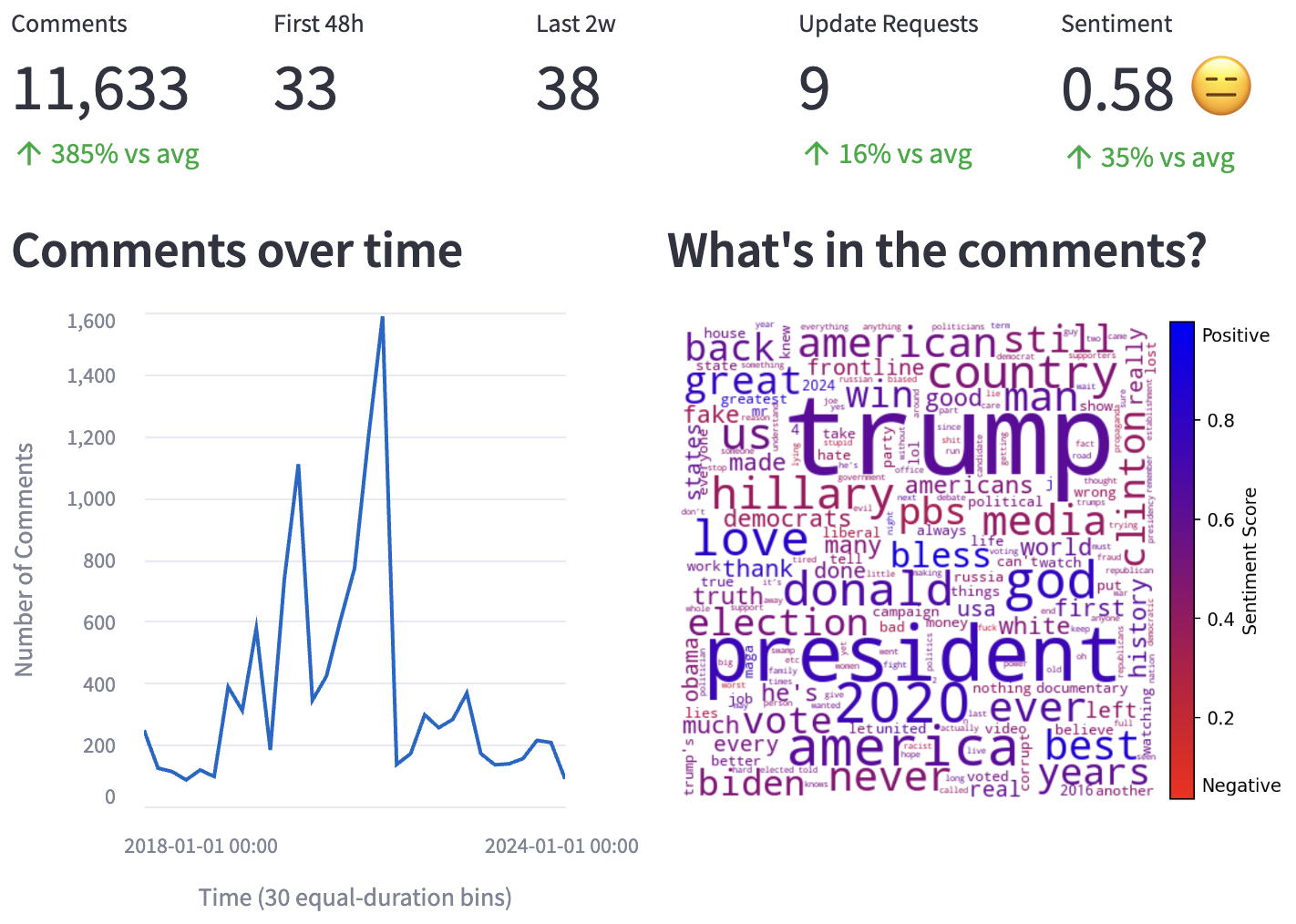}
        \Description{An example of the dashboard view shown for each video.}
        \caption{An example of the dashboard view shown for each video.}
        \label{fig:video-dashboard-view}
    \end{subfigure}
    \hfill 
    \begin{subfigure}[b]{0.45\textwidth}
        \centering
        \includegraphics[width=\textwidth]{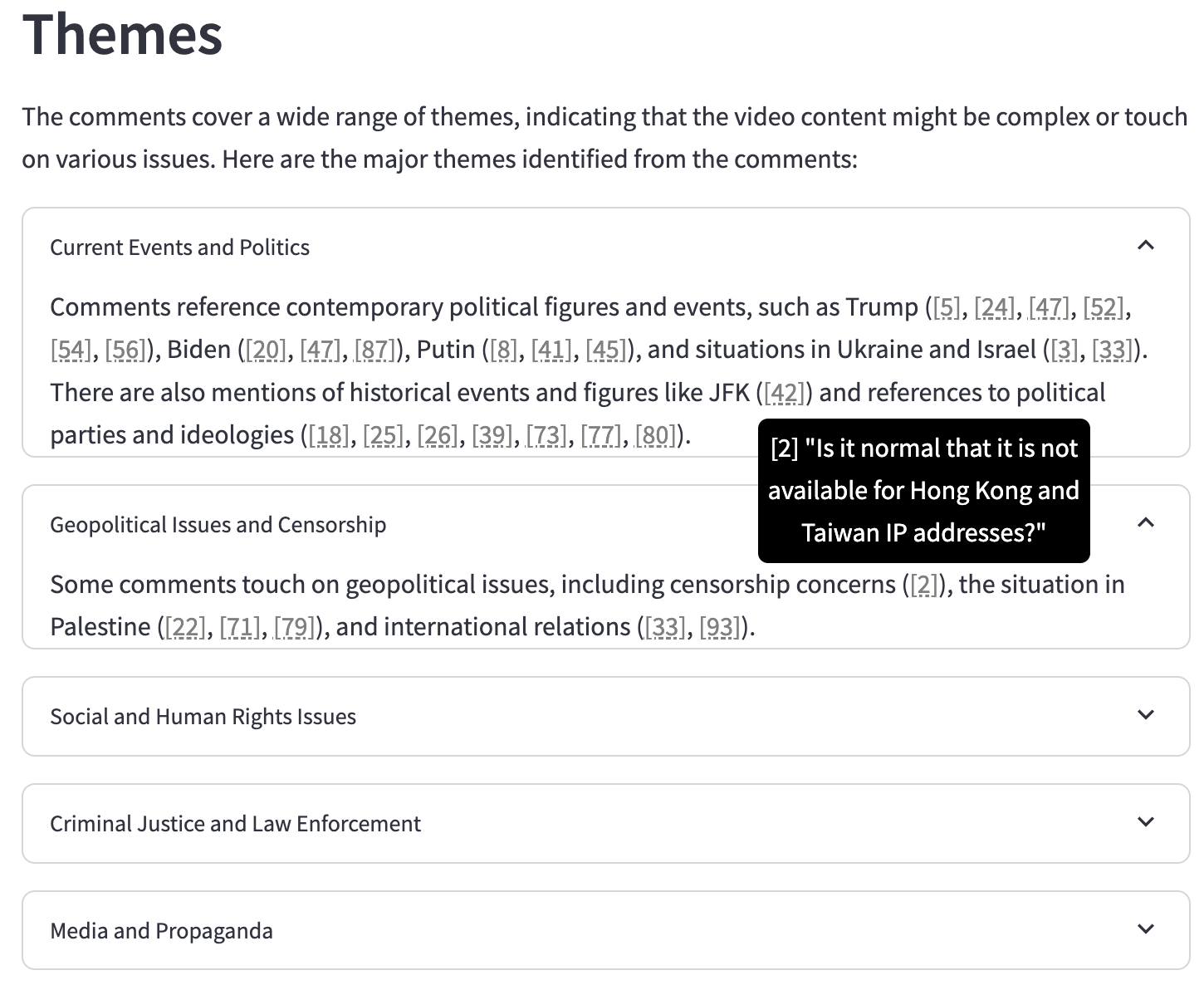}
        \Description{GPT-generated themes at the channel level.}
        \caption{GPT-generated themes at the channel level.}
        \label{fig:channel-themes}
    \end{subfigure}
    \Description{Selected views of the app interface: a video-level dashboard of comments, and channel-level generated themes.}
    \caption{Selected views of the app interface: a video-level dashboard of comments, and channel-level generated themes.}
    \label{fig:interface}
\end{figure}

AudienceView has both backend and frontend components, with the user-facing frontend implemented with the Streamlit framework \cite{streamlitStreamlitFasterWay2021}. In deployment, various aspects can be customized: the YouTube channel to report on, the OpenAI \cite{openaiOpenAIAPI2024} (GPT-4 by default) and local language models \cite{wolfHuggingFaceTransformersStateoftheart2020} used to interpret the comments, the prompts given to downstream language models, and more. It is intended to require limited technical setup and provide an accessible interface for frontend users. At several points in the interface, we make sure to connect generated topics, suggestions, and themes back to the underlying comments (as shown in \autoref{fig:channel-themes}). This is a deliberate design choice intended to give the user more direct engagement with audience feedback, reducing the need to trust an AI system.

\textit{Backend}.
Before frontend use, the system has several backend processes which collect and process YouTube data. Running these steps before deployment is essential to reduce latency for the end user. In addition to collecting comments and videos, these processes calculate summary statistics about videos, create sentiment scores via a local transformer model\footnote{Huggingface: \texttt{cardiffnlp/twitter-roberta-base-sentiment-latest}} \cite{barbieriTweetEvalUnifiedBenchmark2020}, and generate the topic clusters and video- and channel-level themes and suggestions discussed below. 

\subsection{Video tab}
\label{subsec:system-overview-video}
The video tab provides detailed information on each video, shown in \autoref{fig:video-dashboard-view}. It features controls to select a video from the list of all videos in the channel, with options to sort the list chronologically, alphabetically and by measures of engagement. Once a video has been selected, the tab provides summary statistics about its comments, including their average sentiment, and a visualization of their distribution over time. To provide more detailed insight into the comments, there is also a word cloud of the most common terms they contain, color-coded by sentiment score. 

\textit{Comment themes}.
The video tab also offers AI-generated themes detected in the video's comments. These themes are generated by GPT-4 in response to a prompt that includes 100 randomly sampled comments and asks the model to ``summarize the major themes of the comments, and please cite examples.'' The frontend shows the comments the model cites for each theme in tooltips, allowing the user to examine both original content and high-level AI synthesis.

\textit{Suggestions}.
This tab also features GPT-4-generated suggestions for future content in light of the comments, intended as an aid to ideation grounded in audience feedback, rather than a to-do list or ready-made editorial agenda. These suggestions similarly use 100 randomly sampled comments, and ask ``what kinds of new documentary content should [news org] work on?'' As with themes, the prompt instructs the model to cite comments supporting its answers, and the frontend makes them easily accessible in tooltips. (Full prompts are available in the source code.)

\subsection{Channel tab}
The channel tab provides information on the overall YouTube channel, separate from any particular video. The top of this tab features summary statistics like those at the top of the video tab -- among others, the most recent date YouTube data was collected, the number of views, the number of comments, and the average sentiment score of the comments.

\textit{Comment themes and suggestions}.
As on the video tab, the channel tab includes GPT-generated themes and suggestions, using the same prompt and random sampling approach to comments. The comments in this case, however, are sampled from the entire channel rather than only one video, providing a higher-level view of audience interests and feedback.

\begin{wrapfigure}{r}{0.5\textwidth}
    \centering
    \includegraphics[width=0.5\textwidth]{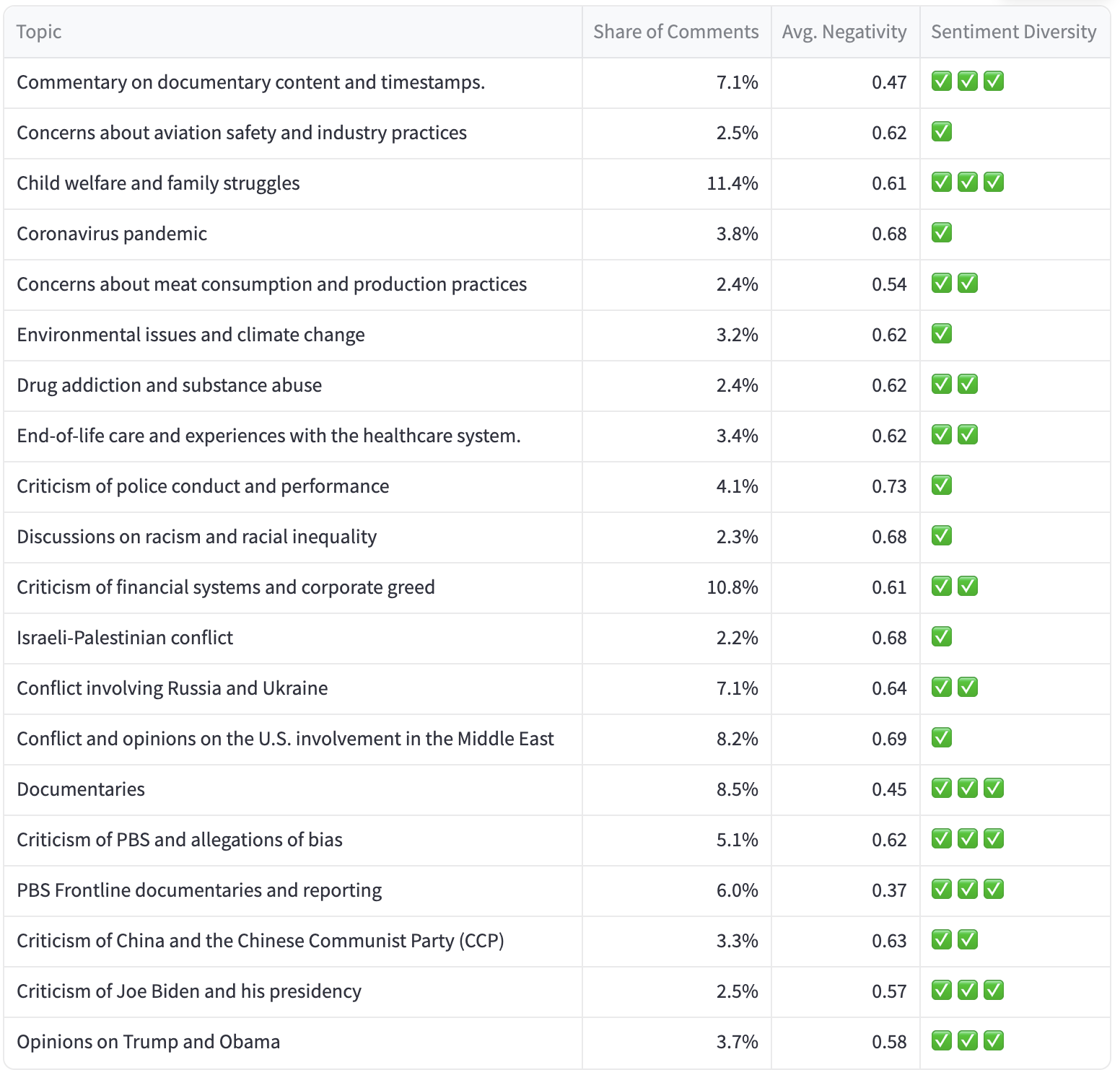}
    \caption{The topics detected in channel-wide comments.}
    \label{fig:channel-topic-table}
\end{wrapfigure}

\textit{Comment topics}.
In addition to themes, we provide an alternative way to explore the topics contained in the comments, using a pipeline that includes all comments rather than only a random sample. These topics are detected by applying HDBSCAN \cite{campelloDensityBasedClusteringBased2013, mcinnesHdbscanHierarchicalDensity2017} clustering to sentence embeddings from the \texttt{all-mpnet-base-v2} sentence-transformers model \cite{reimersSentenceBERTSentenceEmbeddings2019}, after dimensionality reduction with UMAP \cite{bechtDimensionalityReductionVisualizing2019}. GPT-4 is used to label the detected clusters based on a random sample of the comments in each. The frontend displays them in a table (\autoref{fig:channel-topic-table}), with information on the percentage of all comments each cluster accounts for, measures of the variance of sentiment in each cluster, and the ability to browse the constituent comments (not shown in the figure).

\textit{Change alerts}.
AudienceView also flags videos which are experiencing unusually high or low rates of commenting, unusually positive or negative sentiment, or high numbers of comments requesting an updated version of the video. These help journalists detect changes in audience interest quickly. As baselines to compare actual levels to, we use an exponential smoothing model for comments and a simple weighted average by month for sentiment. Requests for updated videos are rare (but informative) enough that a baseline of each video receiving 0 of them is useful.

\textit{``Superfans''}.
The ``superfan'' commenters section indicates the commenters with the highest (i.e., most positive) average sentiment score across all included comments. To avoid bias from commenters with few posts, only those commenters with at least 200 comments are included.

\section{Discussion and Evaluation}
\label{sec:discussion}
The Frontline team at PBS has provided helpful feedback throughout the process of developing AudienceView. We have recently begun conducting structured user interviews; the one interview completed so far had a positive reaction and pointed particularly to the topic detection and change alerts as helpful features. We aim to develop the tool further, focusing especially on two priorities: first, continued refinement of the interface with existing users, under current IRB approval; second, with future approval, more systematic measurement of its efficacy in newsroom workflows. In the future, expansion to modalities or comment sources other than YouTube would also expand the set of potential users.


\bibliographystyle{acmref}
\bibliography{references}


\begin{thebibliography}{24}


\ifx \showCODEN    \undefined \def \showCODEN     #1{\unskip}     \fi
\ifx \showDOI      \undefined \def \showDOI       #1{#1}\fi
\ifx \showISBNx    \undefined \def \showISBNx     #1{\unskip}     \fi
\ifx \showISBNxiii \undefined \def \showISBNxiii  #1{\unskip}     \fi
\ifx \showISSN     \undefined \def \showISSN      #1{\unskip}     \fi
\ifx \showLCCN     \undefined \def \showLCCN      #1{\unskip}     \fi
\ifx \shownote     \undefined \def \shownote      #1{#1}          \fi
\ifx \showarticletitle \undefined \def \showarticletitle #1{#1}   \fi
\ifx \showURL      \undefined \def \showURL       {\relax}        \fi
\providecommand\bibfield[2]{#2}
\providecommand\bibinfo[2]{#2}
\providecommand\natexlab[1]{#1}
\providecommand\showeprint[2][]{arXiv:#2}

\bibitem[Angelov(2020)]%
        {angelovTop2VecDistributedRepresentations2020}
\bibfield{author}{\bibinfo{person}{Dimo Angelov}.} \bibinfo{year}{2020}\natexlab{}.
\newblock \bibinfo{title}{{Top2Vec}: {Distributed} {Representations} of {Topics}}.
\newblock
\newblock
\urldef\tempurl%
\url{http://arxiv.org/abs/2008.09470}
\showURL{%
\tempurl}


\bibitem[Barbieri et~al\mbox{.}(2020)]%
        {barbieriTweetEvalUnifiedBenchmark2020}
\bibfield{author}{\bibinfo{person}{Francesco Barbieri}, \bibinfo{person}{Jose Camacho-Collados}, \bibinfo{person}{Leonardo Neves}, {and} \bibinfo{person}{Luis Espinosa-Anke}.} \bibinfo{year}{2020}\natexlab{}.
\newblock \bibinfo{title}{{TweetEval}: {Unified} {Benchmark} and {Comparative} {Evaluation} for {Tweet} {Classification}}.
\newblock
\newblock
\urldef\tempurl%
\url{http://arxiv.org/abs/2010.12421}
\showURL{%
\tempurl}


\bibitem[Becht et~al\mbox{.}(2019)]%
        {bechtDimensionalityReductionVisualizing2019}
\bibfield{author}{\bibinfo{person}{Etienne Becht}, \bibinfo{person}{Leland McInnes}, \bibinfo{person}{John Healy}, \bibinfo{person}{Charles-Antoine Dutertre}, \bibinfo{person}{Immanuel W~H Kwok}, \bibinfo{person}{Lai~Guan Ng}, \bibinfo{person}{Florent Ginhoux}, {and} \bibinfo{person}{Evan~W Newell}.} \bibinfo{year}{2019}\natexlab{}.
\newblock \showarticletitle{Dimensionality reduction for visualizing single-cell data using {UMAP}}.
\newblock \bibinfo{journal}{\emph{Nature Biotechnology}} \bibinfo{volume}{37}, \bibinfo{number}{1} (\bibinfo{date}{Jan.} \bibinfo{year}{2019}), \bibinfo{pages}{38--44}.
\newblock
\showISSN{1087-0156, 1546-1696}
\urldef\tempurl%
\url{https://doi.org/10/gfkwzq}
\showDOI{\tempurl}


\bibitem[Beeferman and Gillani(2023)]%
        {beefermanFeedbackMapToolMaking2023}
\bibfield{author}{\bibinfo{person}{Doug Beeferman} {and} \bibinfo{person}{Nabeel Gillani}.} \bibinfo{year}{2023}\natexlab{}.
\newblock \showarticletitle{{FeedbackMap}: {A} {Tool} for {Making} {Sense} of {Open}-ended {Survey} {Responses}}. In \bibinfo{booktitle}{\emph{Computer {Supported} {Cooperative} {Work} and {Social} {Computing}}}. \bibinfo{publisher}{ACM}, \bibinfo{address}{Minneapolis MN USA}, \bibinfo{pages}{395--397}.
\newblock
\showISBNx{9798400701290}
\urldef\tempurl%
\url{https://doi.org/10/mvng}
\showDOI{\tempurl}


\bibitem[Blei et~al\mbox{.}(2003)]%
        {bleiLatentDirichletAllocation2003}
\bibfield{author}{\bibinfo{person}{David~M. Blei}, \bibinfo{person}{Andrew~Y. Ng}, {and} \bibinfo{person}{Michael~I. Jordan}.} \bibinfo{year}{2003}\natexlab{}.
\newblock \showarticletitle{Latent dirichlet allocation}.
\newblock \bibinfo{journal}{\emph{Journal of Machine Learning Research}}  \bibinfo{volume}{3} (\bibinfo{year}{2003}), \bibinfo{pages}{993--1022}.
\newblock
\showISSN{1532-4435}


\bibitem[Campello et~al\mbox{.}(2013)]%
        {campelloDensityBasedClusteringBased2013}
\bibfield{author}{\bibinfo{person}{Ricardo J. G.~B. Campello}, \bibinfo{person}{Davoud Moulavi}, {and} \bibinfo{person}{Joerg Sander}.} \bibinfo{year}{2013}\natexlab{}.
\newblock \showarticletitle{Density-{Based} {Clustering} {Based} on {Hierarchical} {Density} {Estimates}}.
\newblock In \bibinfo{booktitle}{\emph{Advances in {Knowledge} {Discovery} and {Data} {Mining}}}, \bibfield{editor}{\bibinfo{person}{David Hutchison}, \bibinfo{person}{Takeo Kanade}, \bibinfo{person}{Josef Kittler}, \bibinfo{person}{Jon~M. Kleinberg}, \bibinfo{person}{Friedemann Mattern}, \bibinfo{person}{John~C. Mitchell}, \bibinfo{person}{Moni Naor}, \bibinfo{person}{Oscar Nierstrasz}, \bibinfo{person}{C.~Pandu~Rangan}, \bibinfo{person}{Bernhard Steffen}, \bibinfo{person}{Madhu Sudan}, \bibinfo{person}{Demetri Terzopoulos}, \bibinfo{person}{Doug Tygar}, \bibinfo{person}{Moshe~Y. Vardi}, \bibinfo{person}{Gerhard Weikum}, \bibinfo{person}{Jian Pei}, \bibinfo{person}{Vincent~S. Tseng}, \bibinfo{person}{Longbing Cao}, \bibinfo{person}{Hiroshi Motoda}, {and} \bibinfo{person}{Guandong Xu}} (Eds.). Vol.~\bibinfo{volume}{7819}. \bibinfo{publisher}{Springer Berlin Heidelberg}, \bibinfo{address}{Berlin, Heidelberg}, \bibinfo{pages}{160--172}.
\newblock
\showISBNx{978-3-642-37455-5 978-3-642-37456-2}
\urldef\tempurl%
\url{https://doi.org/10.1007/978-3-642-37456-2_14}
\showDOI{\tempurl}
\newblock
\shownote{Series Title: Lecture Notes in Computer Science}.


\bibitem[{Chartbeat, Inc.}(2024)]%
        {chartbeatinc.Chartbeat2024}
\bibfield{author}{\bibinfo{person}{{Chartbeat, Inc.}}} \bibinfo{year}{2024}\natexlab{}.
\newblock \bibinfo{title}{Chartbeat}.
\newblock
\newblock
\urldef\tempurl%
\url{https://chartbeat.com/}
\showURL{%
\tempurl}


\bibitem[Goldman et~al\mbox{.}(2022)]%
        {goldmanQuADDeepLearningAssisted2022}
\bibfield{author}{\bibinfo{person}{Ariel Goldman}, \bibinfo{person}{Cindy Espinosa}, \bibinfo{person}{Shivani Patel}, \bibinfo{person}{Francesca Cavuoti}, \bibinfo{person}{Jade Chen}, \bibinfo{person}{Alexandra Cheng}, \bibinfo{person}{Sabrina Meng}, \bibinfo{person}{Aditi Patil}, \bibinfo{person}{Lydia~B Chilton}, {and} \bibinfo{person}{Sarah Morrison-Smith}.} \bibinfo{year}{2022}\natexlab{}.
\newblock \showarticletitle{{QuAD}: {Deep}-{Learning} {Assisted} {Qualitative} {Data} {Analysis} with {Affinity} {Diagrams}}. In \bibinfo{booktitle}{\emph{{CHI} {Conference} on {Human} {Factors} in {Computing} {Systems} {Extended} {Abstracts}}}. \bibinfo{publisher}{ACM}, \bibinfo{address}{New Orleans LA USA}, \bibinfo{pages}{1--7}.
\newblock
\showISBNx{978-1-4503-9156-6}
\urldef\tempurl%
\url{https://doi.org/10/gtt393}
\showDOI{\tempurl}


\bibitem[Grootendorst(2022)]%
        {grootendorstBERTopicNeuralTopic2022}
\bibfield{author}{\bibinfo{person}{Maarten Grootendorst}.} \bibinfo{year}{2022}\natexlab{}.
\newblock \bibinfo{title}{{BERTopic}: {Neural} topic modeling with a class-based {TF}-{IDF} procedure}.
\newblock
\newblock
\urldef\tempurl%
\url{https://arxiv.org/abs/2203.05794}
\showURL{%
\tempurl}


\bibitem[Hermida(2012)]%
        {hermidaSocialJournalismExploring2012}
\bibfield{author}{\bibinfo{person}{Alfred Hermida}.} \bibinfo{year}{2012}\natexlab{}.
\newblock \showarticletitle{Social {Journalism}: {Exploring} how {Social} {Media} is {Shaping} {Journalism}}.
\newblock In \bibinfo{booktitle}{\emph{The {Handbook} of {Global} {Online} {Journalism}} (\bibinfo{edition}{1st} ed.)}, \bibfield{editor}{\bibinfo{person}{Eugenia Siapera} {and} \bibinfo{person}{Andreas Veglis}} (Eds.). \bibinfo{publisher}{Wiley}, \bibinfo{address}{Chichester, UK}, \bibinfo{pages}{309--328}.
\newblock
\showISBNx{978-1-4443-3855-3}
\urldef\tempurl%
\url{https://doi.org/10/ks59}
\showDOI{\tempurl}


\bibitem[Lawrence(2020)]%
        {lawrenceCampaignNewsTime2020}
\bibfield{author}{\bibinfo{person}{Regina Lawrence}.} \bibinfo{year}{2020}\natexlab{}.
\newblock \showarticletitle{Campaign {News} in the {Time} of {Twitter}}.
\newblock In \bibinfo{booktitle}{\emph{Controlling the {Message}}}, \bibfield{editor}{\bibinfo{person}{V.~Farrar-Myers} {and} \bibinfo{person}{J.~Vaughn}} (Eds.). \bibinfo{publisher}{NYU Press}, \bibinfo{address}{New York}, \bibinfo{pages}{93--112}.
\newblock
\showISBNx{978-1-4798-6550-5}
\urldef\tempurl%
\url{https://doi.org/10/jr3n}
\showDOI{\tempurl}


\bibitem[{Lumivero}(2023)]%
        {lumiveroNVivo2023}
\bibfield{author}{\bibinfo{person}{{Lumivero}}.} \bibinfo{year}{2023}\natexlab{}.
\newblock \bibinfo{title}{{NVivo}}.
\newblock
\newblock
\urldef\tempurl%
\url{https://lumivero.com/products/nvivo/}
\showURL{%
\tempurl}


\bibitem[Marwick and {danah boyd}(2011)]%
        {marwickTweetHonestlyTweet2011}
\bibfield{author}{\bibinfo{person}{Alice~E. Marwick} {and} \bibinfo{person}{{danah boyd}}.} \bibinfo{year}{2011}\natexlab{}.
\newblock \showarticletitle{I tweet honestly, {I} tweet passionately: {Twitter} users, context collapse, and the imagined audience}.
\newblock \bibinfo{journal}{\emph{New Media \& Society}} \bibinfo{volume}{13}, \bibinfo{number}{1} (\bibinfo{date}{Feb.} \bibinfo{year}{2011}), \bibinfo{pages}{114--133}.
\newblock
\showISSN{1461-4448, 1461-7315}
\urldef\tempurl%
\url{https://doi.org/10/ffn6sf}
\showDOI{\tempurl}


\bibitem[McInnes et~al\mbox{.}(2017)]%
        {mcinnesHdbscanHierarchicalDensity2017}
\bibfield{author}{\bibinfo{person}{Leland McInnes}, \bibinfo{person}{John Healy}, {and} \bibinfo{person}{Steve Astels}.} \bibinfo{year}{2017}\natexlab{}.
\newblock \showarticletitle{hdbscan: {Hierarchical} density based clustering}.
\newblock \bibinfo{journal}{\emph{JOSS}} \bibinfo{volume}{2}, \bibinfo{number}{11} (\bibinfo{year}{2017}), \bibinfo{pages}{205}.
\newblock
\showISSN{2475-9066}
\urldef\tempurl%
\url{https://doi.org/10/ggfp85}
\showDOI{\tempurl}


\bibitem[{OpenAI}(2024)]%
        {openaiOpenAIAPI2024}
\bibfield{author}{\bibinfo{person}{{OpenAI}}.} \bibinfo{year}{2024}\natexlab{}.
\newblock \bibinfo{title}{{OpenAI} {API}}.
\newblock
\newblock
\urldef\tempurl%
\url{https://platform.openai.com/}
\showURL{%
\tempurl}


\bibitem[Overney et~al\mbox{.}(2024)]%
        {overneySenseMateAccessibleBeginnerFriendly2024}
\bibfield{author}{\bibinfo{person}{Cassandra Overney}, \bibinfo{person}{Belén Saldías}, \bibinfo{person}{Dimitra Dimitrakopoulou}, {and} \bibinfo{person}{Deb Roy}.} \bibinfo{year}{2024}\natexlab{}.
\newblock \showarticletitle{{SenseMate}: {An} {Accessible} and {Beginner}-{Friendly} {Human}-{AI} {Platform} for {Qualitative} {Data} {Analysis}}. In \bibinfo{booktitle}{\emph{Proceedings of {IUI} '24}}. \bibinfo{publisher}{ACM}, \bibinfo{address}{Greenville}, \bibinfo{pages}{922--939}.
\newblock
\showISBNx{9798400705083}
\urldef\tempurl%
\url{https://doi.org/10/gtt392}
\showDOI{\tempurl}


\bibitem[Pavlik(2023)]%
        {pavlikCollaboratingChatGPTConsidering2023}
\bibfield{author}{\bibinfo{person}{John~V. Pavlik}.} \bibinfo{year}{2023}\natexlab{}.
\newblock \showarticletitle{Collaborating {With} {ChatGPT}: {Considering} the {Implications} of {Generative} {Artificial} {Intelligence} for {Journalism} and {Media} {Education}}.
\newblock \bibinfo{journal}{\emph{Journalism \& Mass Communication Educator}} \bibinfo{volume}{78}, \bibinfo{number}{1} (\bibinfo{year}{2023}), \bibinfo{pages}{84--93}.
\newblock
\showISSN{1077-6958, 2161-4326}
\urldef\tempurl%
\url{https://doi.org/10/js2j}
\showDOI{\tempurl}


\bibitem[Reimers and Gurevych(2019)]%
        {reimersSentenceBERTSentenceEmbeddings2019}
\bibfield{author}{\bibinfo{person}{Nils Reimers} {and} \bibinfo{person}{Iryna Gurevych}.} \bibinfo{year}{2019}\natexlab{}.
\newblock \bibinfo{title}{Sentence-{BERT}: {Sentence} {Embeddings} using {Siamese} {BERT}-{Networks}}.
\newblock
\newblock
\urldef\tempurl%
\url{http://arxiv.org/abs/1908.10084}
\showURL{%
\tempurl}


\bibitem[Robinson(2019)]%
        {robinsonAudienceMindEye2019}
\bibfield{author}{\bibinfo{person}{James~G. Robinson}.} \bibinfo{year}{2019}\natexlab{}.
\newblock \showarticletitle{The {Audience} in the {Mind}'s {Eye}: {How} {Journalists} {Imagine} {Their} {Readers}}.
\newblock \bibinfo{journal}{\emph{Columbia J. Rev.}} (\bibinfo{year}{2019}).
\newblock
\urldef\tempurl%
\url{https://doi.org/10/gtt9sx}
\showDOI{\tempurl}


\bibitem[Stray(2019)]%
        {strayMakingArtificialIntelligence2019}
\bibfield{author}{\bibinfo{person}{Jonathan Stray}.} \bibinfo{year}{2019}\natexlab{}.
\newblock \showarticletitle{Making {Artificial} {Intelligence} {Work} for {Investigative} {Journalism}}.
\newblock \bibinfo{journal}{\emph{Digital Journalism}} \bibinfo{volume}{7}, \bibinfo{number}{8} (\bibinfo{year}{2019}), \bibinfo{pages}{1076--1097}.
\newblock
\showISSN{2167-0811, 2167-082X}
\urldef\tempurl%
\url{https://doi.org/10/gf4j6r}
\showDOI{\tempurl}


\bibitem[{Streamlit}(2021)]%
        {streamlitStreamlitFasterWay2021}
\bibfield{author}{\bibinfo{person}{{Streamlit}}.} \bibinfo{year}{2021}\natexlab{}.
\newblock \bibinfo{title}{Streamlit -- {A} faster way to build and share data apps}.
\newblock
\newblock
\urldef\tempurl%
\url{https://streamlit.io/}
\showURL{%
\tempurl}


\bibitem[{VERBI Software}(2022)]%
        {verbisoftwareMAXQDA2022}
\bibfield{author}{\bibinfo{person}{{VERBI Software}}.} \bibinfo{year}{2022}\natexlab{}.
\newblock \bibinfo{title}{{MAXQDA}}.
\newblock
\newblock
\urldef\tempurl%
\url{https://www.maxqda.com/}
\showURL{%
\tempurl}


\bibitem[Whipple and Shermak(2018)]%
        {whippleQualityQuantityPolicy2018}
\bibfield{author}{\bibinfo{person}{Kelsey Whipple} {and} \bibinfo{person}{Jeremy Shermak}.} \bibinfo{year}{2018}\natexlab{}.
\newblock \showarticletitle{Quality, quantity and policy: {How} newspaper journalists use digital metrics to evaluate their performance and their papers’ strategies}.
\newblock \bibinfo{journal}{\emph{\#ISOJ Journal}} \bibinfo{volume}{8}, \bibinfo{number}{1} (\bibinfo{year}{2018}), \bibinfo{pages}{67--88}.
\newblock
\showISSN{2328-0700}


\bibitem[Wolf et~al\mbox{.}(2020)]%
        {wolfHuggingFaceTransformersStateoftheart2020}
\bibfield{author}{\bibinfo{person}{Thomas Wolf}, \bibinfo{person}{Lysandre Debut}, \bibinfo{person}{Victor Sanh}, \bibinfo{person}{Julien Chaumond}, \bibinfo{person}{Clement Delangue}, \bibinfo{person}{Anthony Moi}, \bibinfo{person}{Pierric Cistac}, \bibinfo{person}{Tim Rault}, \bibinfo{person}{Rémi Louf}, \bibinfo{person}{Morgan Funtowicz}, \bibinfo{person}{Joe Davison}, \bibinfo{person}{Sam Shleifer}, \bibinfo{person}{Patrick von Platen}, \bibinfo{person}{Clara Ma}, \bibinfo{person}{Yacine Jernite}, \bibinfo{person}{Julien Plu}, \bibinfo{person}{Canwen Xu}, \bibinfo{person}{Teven~Le Scao}, \bibinfo{person}{Sylvain Gugger}, \bibinfo{person}{Mariama Drame}, \bibinfo{person}{Quentin Lhoest}, {and} \bibinfo{person}{Alexander~M. Rush}.} \bibinfo{year}{2020}\natexlab{}.
\newblock \bibinfo{title}{{HuggingFace}'s {Transformers}: {State}-of-the-art {Natural} {Language} {Processing}}.
\newblock
\newblock
\urldef\tempurl%
\url{http://arxiv.org/abs/1910.03771}
\showURL{%
\tempurl}


\end{thebibliography}

\end{document}